\documentclass[english]{sbrt}
\usepackage[english]{babel}
\usepackage[utf8]{inputenc}

\usepackage{booktabs}
\usepackage{graphicx}
\usepackage{amsfonts}
\usepackage{stfloats}
\usepackage{cite}
\usepackage{balance}
\usepackage{float}

\begin{document}

\title{ONLINE ML-BASED JOINT CHANNEL ESTIMATION AND MIMO DECODING FOR DYNAMIC CHANNELS}

\author{Name1 Surname1 and Name2 Surname2
\thanks{Name1 Surname1, Department1, University1, City1-State1, e-mail: xxxxx@yyyyy.zzzzz.br; Name2 Surname2, Department2, University2, City2-state2, e-mail: xxxxx@yyyyy.zzzzz.br. This work was partially supported by XXXXXXX (XX/XXXXX-X).}%
}

\author{Luiz Fernando Moreira Teixeira, Vinicius Henrique Luiz, Jonathan Aguiar Soares, \\Kayol Soares Mayer, and Dalton Soares Arantes
\thanks{Luiz Fernando M. Teixeira, Vinicius H. Luiz, Jonathan A. Soares, Kayol S. Mayer, and Dalton S. Arantes are with the Digital Communications Laboratory -- ComLab, Department of Communications, School of Electrical and Computer Engineering, University of
Campinas -- Unicamp, 13083-852 Campinas, SP, Brazil, e-mails: l182735@dac.unicamp.br; v245342@dac.unicamp.br; j229966@dac.unicamp.br; kayol@unicamp.br; dalton@unicamp.br.}
}

\maketitle

\markboth{XLII BRAZILIAN SYMPOSIUM ON TELECOMMUNICATIONS AND SIGNAL PROCESSING - SBrT 2024, OCTOBER 01--04, 2024, BELÉM, PA}{}


\begin{abstract}
This paper presents an online method for joint channel estimation and decoding in massive MIMO-OFDM systems using complex-valued neural networks (CVNNs). The study evaluates the performance of various CVNNs, such as the complex-valued feedforward neural network (CVFNN), split-complex feedforward neural network (SCFNN), complex radial basis function (C-RBF), fully-complex radial basis function (FC-RBF) and phase-transmittance radial basis function (PT-RBF), in realistic 5G communication scenarios. Results demonstrate improvements in mean squared error (MSE), convergence, and bit error rate (BER) accuracy. The C-RBF and PT-RBF architectures show the most promising outcomes, suggesting that RBF-based CVNNs provide a reliable and efficient solution for complex and noisy communication environments. These findings have potential implications for applying advanced neural network techniques in next-generation wireless systems.
\end{abstract}


\section{Introduction}
In recent years, massive MIMO technology has become increasingly important in wireless communications due to its enhancements in system capacity and energy efficiency \cite{Hua2022}. Among the various methods employed in massive MIMO, spatial diversity plays a crucial role, especially when information about the channel is unknown at the transmitter. Space-time block coding (STBC) is one such method that enables the transmission of orthogonal or quasi-orthogonal signals amidst varying fading conditions across multiple antennas \cite{Morsali2019, Asim2022}.

Additionally, the integration of orthogonal frequency-division multiplexing (OFDM) with STBC is commonplace, facilitating efficient data transmission to multiple users across closely spaced subchannels without the need for extensive channel equalization \cite{Chen2017, Chen2019, Soares2021}.

While numerous studies have addressed joint channel estimation and decoding in massive MIMO systems \cite{Wu2016, Verenzuela2020, Sanoopkumar2022}, few have specifically tackled the challenges associated with quasi-orthogonal STBC (QOSTBC) in massive MIMO-OFDM configurations. This complexity arises from the difficulty in creating quasi-orthogonal matrices and decoding algorithms for a massive number of antennas and $M$-ary quadrature amplitude modulation ($M$-QAM). In the literature, the MIMO phase-transmittance radial basis function (PT-RBF) neural network proposed by Soares et al.~\cite{Soares2021} seems to be the only work addressing this issue.

The PT-RBF belongs to the class of complex-valued neural networks (CVNNs)~\cite{mayer2022b}. Unlike real-valued neural networks~(RVNNs), CVNNs are able to handle complex inputs and outputs directly. Therefore, CVNNs should be a natural choice for complex-valued signals and even be studied for real-valued applications. For example, considering the XOR problem, derived from the ``AND/OR'' theorem of Minsky and Papert \cite{Minsky1969} for two dimensions, a single real-valued perceptron cannot learn the XOR function. As stated by Rumelhart and McClelland \cite{Rumelhart1986}, at least a three-layered RVNN is necessary to solve the XOR problem. On the other hand, only a single complex-valued neuron is needed to circumvent Minsky and Papert's limitation \cite{Nitta2003}. Nevertheless, the use of a single complex-valued neuron is not the sole motivation, since with CVNN architectures it is possible to increase the functionality of neural networks, improving their performance and reducing the training time compared with RVNNs \cite{Hirose2012, Zhang2022}.

In the context of RVNNs applied to complex-valued problems, two trivial dual univariate solutions are dual-univariate and split-input RVNN architectures. Dual-univariate RVNNs are simple due to their straightforward use. However, as dual-univariate RVNNs split the complex-valued input into their real and imaginary components and process them separately using two real-valued neural networks, they are only suitable for phase-independent systems. Conversely, as split-input RVNNs deal with real and imaginary components with one unique neural network, they are able to work with phase-dependent systems. Nonetheless, split-input RVNNs have some phase-recovery issues due to the real and imaginary components split at the RVNN input and output \cite{Hirose2012b}.

In such a context, this paper compiles the CVNNs results for realistic massive MIMO-OFDM communications, available on the Ph.D. Thesis ``Complex-Valued Neural Networks and Applications in Telecommunications'', proposed by Mayer~\cite{mayer2022b}. In this work, we adopt the same massive-MIMO architecture described by Soares et al.~\cite{Soares2021} to validate the effectiveness of CVNNs for joint channel estimation and decoding. The CVNNs considered in this work are the complex-valued feedforward neural network (CVFNN), split-complex feedforward neural network (SCFNN), complex-valued radial basis function (C-RBF), fully complex-valued radial basis function (FC-RBF), and PT-RBF.

\begin{figure*}[b]
\centering
\includegraphics[width=\linewidth]{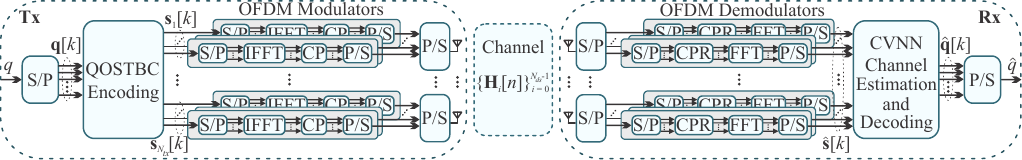}
\caption{Massive MIMO-OFDM architecture with QOSTBC spatial diversity and $N_{tx}$ and $N_{rx}$ transmitting and receiving antennas, respectively.}
\label{fig:mimo_arc}
\end{figure*}

\section{System Model}

\subsection{Complex-valued Neural Networks}

In counterpart of RVNNs applied to complex-valued problems, several CVNNs have been proposed in the past three decades \cite{Chen1994,Dong2021,Dramsch2021,Enriconi2020,Freire2020,Kim2002,Loss2007,Mayer2019a,Mayer2020a,Mayer10361866,Savitha2009,Savitha2012,Zhang2019}. One of the most studied CVNNs in the literature is the CVFNN, a multilayer perceptron without feedback among layers, adapted to directly process data in the complex domain \cite{Dong2021,Freire2020,Kim2002,Zhang2019}. CVFNNs can operate with fully complex transcendental activation functions that satisfy the Cauchy-Riemann equations \cite{Utreras2019} with relaxed conditions, such as circular, inverse circular, hyperbolic, and inverse hyperbolic functions \cite{Kim2002}. Also, an important and particular case of CVFNNs is the SCFNN, in which real and imaginary components are separately processed by holomorphic functions (i.e., analytic functions) in $\mathbb{R}$ \cite{Kim2002,Scardapane2020}. Based on a different CVNN architecture, the C-RBF neural network can also operate with complex numbers \cite{Chen1994}. Notwithstanding, due to the C-RBF phase vanishing under the Euclidean norm from Gaussian neurons, Loss et al. \cite{Loss2007} proposed the phase transmittance radial basis function (PT-RBF) neural network, which considers split-complex Gaussian neurons to circumvent any phase issue \cite{Enriconi2020,Loss2007,Mayer2019a,Mayer2020a}. Also, taking that into account, Savitha et al. \cite{Savitha2009} proposed the FC-RBF neural network, in which $\mathrm{sech}(\cdot)$ activation functions map $\mathbb{C}^{N}\mapsto\mathbb{C}$ with Gaussian-like characteristics \cite{Savitha2009,Savitha2012}. 

The mathematical background of the CVNNs discussed in this work can be found in Section 3 of~\cite{mayer2022b}, including the forward and backward equations utilized for training and inference. Section 4 of~\cite{mayer2022b} also provides a detailed computational complexity analysis.

\subsection{System Architecture}
Fig.~\ref{fig:mimo_arc} depicts the architecture of the massive MIMO-OFDM system with QOSTBC (quasi-orthogonal space-time block code) spatial diversity. In this setup, \(N_{tx}\) and \(N_{rx}\) represent the number of transmitting and receiving antennas, respectively. On the transmitter side, QAM symbols \(q\) are parallelized into the vector \(\mathbf{q}[k] \in \mathbb{C}^{N_{s}}\) using a serial-to-parallel (S/P) block. This vector is then encoded spatially and temporally by the QOSTBC encoding block, forming the QOSTBC matrix \(\mathbf{S}[k] \in \mathbb{C}^{N_{tp} \times N_{tx}}\). Each column \(\mathbf{s}_{n_{tx}}[k] \in \mathbb{C}^{N_{tp}}\) of \(\mathbf{S}[k]\) feeds into the OFDM (orthogonal frequency-division multiplexing) modulators, where the signal is converted from the frequency domain to the time domain using an inverse fast Fourier transform (IFFT), and a cyclic prefix (CP) is added to prevent inter-symbol interference (ISI).

Considering a sample-spaced multipath channel with \(N_{ds}\) samples \(\left\{\mathbf{H}_i[n]\right\}_{i=0}^{N_{ds}-1} \in \mathbb{C}^{N_{rx} \times N_{tx}}\), the received signal is:
\[
\mathbf{y}[n] = \sum_{i=0}^{N_{ds}-1}\mathbf{H}_i[n]\mathbf{x}[n-i] + \mathbf{w}[n],
\]
where \(\mathbf{x}[n] \in \mathbb{C}^{N_{tx}}\) is the transmitted data vector and \(\mathbf{w}[n] \sim \mathcal{C}\mathcal{N}(0,\,\sigma_w^{2}) \in \mathbb{C}^{N_{rx}}\) represents the complex additive white Gaussian noise (AWGN) at the receiver, with zero mean and variance \(\sigma_w^{2}\).

At the receiver, the signal is parallelized and fed into OFDM demodulators. After removing the cyclic prefix and converting it back to the frequency domain using a fast Fourier transform (FFT), the signal is serialized. The received QOSTBC vector \(\mathbf{\hat{s}}[k] \in \mathbb{C}^{N_{tp} \times N_{rx}}\) is then input to the CVNN for channel estimation and decoding, producing the estimated output \(\mathbf{\hat{q}}[k] \in \mathbb{C}^{N_{s}}\).

This CVNN-based architecture facilitates efficient channel decoding of QOSTBC in massive MIMO-OFDM systems with high-order \(M\)-QAM, offering flexibility and competitive computational complexity as shown by Soares et al.~\cite{Soares2021}. For more details, see \cite{10233097} Section II-B.

\section{results}
To illustrate a practical scenario, we configured the simulation system according to the 3GPP TS 38.211 specification for 5G physical channels and modulation \cite{3gpp.38.211}. The OFDM setup includes a 60 kHz subcarrier spacing, 256 active subcarriers, and a block-based pilot scheme with a sampling rate of 1/6. The symbols are modulated with 16-QAM. For the massive MIMO configuration, we employed 32 antennas at both the transmitter and receiver, i.e., $N_{tx}=N_{rx}=32$.

Based on the tapped delay line-A (TDL-A) model specified in the 3GPP TR 38.901 5G channel models~\cite{3gpp.38.901}, the massive MIMO channel is described by the TDLA30-5 model, outlined in the 3GPP TR 38.104 document on 5G base station radio transmission and reception~\cite{3gpp.38.104}. The TDLA30-5 model is characterized by 12 taps, with delays ranging from 0.0 ns to 290 ns and power levels spanning from -26.2 dB to 0 dB. A Rayleigh distribution is employed to calculate each sub-channel within the set $\left\{\mathbf{H}_i[n]\right\}_{i=0}^{N_{ds}-1}$ (see Table G.2.1.2-2 in \cite{3gpp.38.104}), incorporating a maximum Doppler frequency of 5 Hz to effectively simulate the channel's dynamics (see Table G.2.2-1 in \cite{3gpp.38.104}). 

The channel estimators and massive MIMO decoders operate with 1024 inputs and 32 outputs. Inputs are derived from the pilots of the OFDM demodulator outputs, taken one at a time, i.e., $\mathbf{\hat{s}}[k]$ for $k\in \left[ 1, 2, \dots, 256\right]$. Each pilot block corresponds to a desired output vector for the $k$-th subcarrier, labeled as $\mathbf{q}[k]$. To maintain real-time system compatibility, each pilot block undergoes processing only once, corresponding to a single training epoch per pilot block. However, to address the Doppler shift, a training upsampling of thirty times, without shuffle, was implemented. Thirty iterations were required to stabilize the learning curve for convergence of the CVNNs. The CVFNN and SCFNN architectures consist of two layers of neurons. Given the necessity of a hundred neurons for joint channel estimation and decoding \cite{Soares2021}, the PT-RBF was designed with a shallow architecture to minimize computational complexity and mitigate convergence issues. The CVFNN and SCFNN employ $\mathrm{arctanh}(\cdot)$ and $\mathrm{tanh}(\cdot)$ activation functions in the hidden layer, respectively, and linear activation functions in the output layer. In the PT-RBF, to ensure kernel stability, the real and imaginary parts of each variance component are bounded by $\epsilon>0$. Dropout was not incorporated as overfitting was not observed in the CVNNs for this application. The hyperparameters of the CVNNs, detailed in Table~\ref{tab:cvnns_hyper_dec}, were empirically determined through trial and error.
\begin{table*}[htbp]
\centering
\caption{CVNN hyperparameters for joint channel estimation and decoding of massive MIMO communication systems.}%
\label{tab:cvnns_hyper_dec}
\footnotesize
\begin{tabular}{lcccccccccc}
\toprule
\textbf{CVNN}& $\eta_w$ & $\eta_b$ & $\eta_\gamma$ & $\eta_\sigma$ & $\eta_\upsilon$ & $\alpha$ & $\mu[0]$ & $\lambda$ & $I^{\{1\}}$ & $I^{\{2\}}$\\
\midrule \midrule
CVFNN  & $0.0125$ & $0.0125$ & $\mathbf{-}$ & $\mathbf{-}$ & $\mathbf{-}$ & $0.00125$ & $0.0100$ & $20$ & \phantom{x}$68$ & $32$\\[0.3cm]
SCFNN  & $0.0125$ & $0.0125$ & $\mathbf{-}$ & $\mathbf{-}$ & $\mathbf{-}$ & $0.00125$ & $0.0100$ & $20$ & \phantom{x}$68$ & $32$\\[0.3cm]
C-RBF  & $0.0100$ & $0.0200$ & $0.0200$ & $0.0100$ & $\mathbf{-}$ & $0.00100$ & $0.0100$ & 20 & $100$ & $\mathbf{-}$\\[0.3cm]
FC-RBF & $0.0125$ & $\mathbf{-}$ & $0.0125$ & $\mathbf{-}$ & $0.0125$ & $0.00125$ & $0.0100$ & $20$ & $100$ & $\mathbf{-}$\\[0.3cm]
PT-RBF & $0.0020$ & $0.0200$ & $0.0200$ & $0.0100$ & $\mathbf{-}$ & $0.00020$ & $0.0100$ & $20$ & $100$ & $\mathbf{-}$\\
\bottomrule
\end{tabular}%
\end{table*}

\begin{figure}[b]
\includegraphics[width=\linewidth]{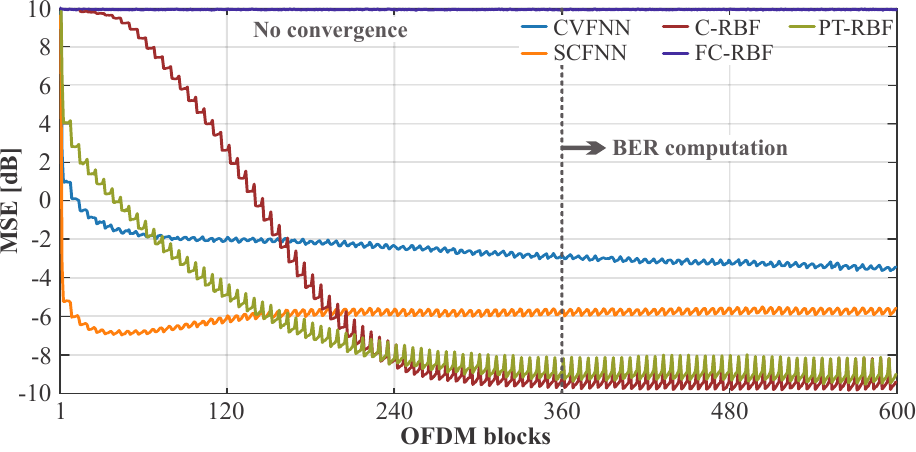}
\caption{MSE convergence results of training and inference of the CVNNs for joint massive MIMO channel estimation and decoding with an $E_b/N_0 = 20$~dB. Results were averaged over ten subsequent simulations.}
\label{fig:mimo_mse}
\end{figure}

Fig.~\ref{fig:mimo_mse} demonstrates the evolution of the mean squared error (MSE) for the CVNNs in the context of joint channel estimation and decoding in massive MIMO communication systems at an energy per bit to noise power spectral density ratio $E_b/N_0 = 20$~dB. Each MSE curve is derived from the average of ten consecutive simulations. As depicted in Fig.~\ref{fig:mimo_mse}, both training and inference phases exhibit a cyclic MSE pattern due to the pilot sampling rate of 1/6. The gradual increase in MSE is attributed to the Doppler effect, while the decrease corresponds to training adjustments in response to pilot blocks. Among the various CVNNs, the FC-RBF was the only one that failed to converge, maintaining an MSE of approximately 10 dB. This can be attributed to the complexity of the problem and the relatively low $E_b/N_0$. Conversely, the C-RBF showed the best convergence, achieving a steady-state $\mathrm{MSE}\approx -9.2$~dB. The CVFNN, SCFNN, and PT-RBF achieved a steady-state MSE of approximately $-3.4$~dB, $-5.6$~dB, and $-9.0$~dB, respectively. 

\begin{figure}[b]
\includegraphics[width=\linewidth]{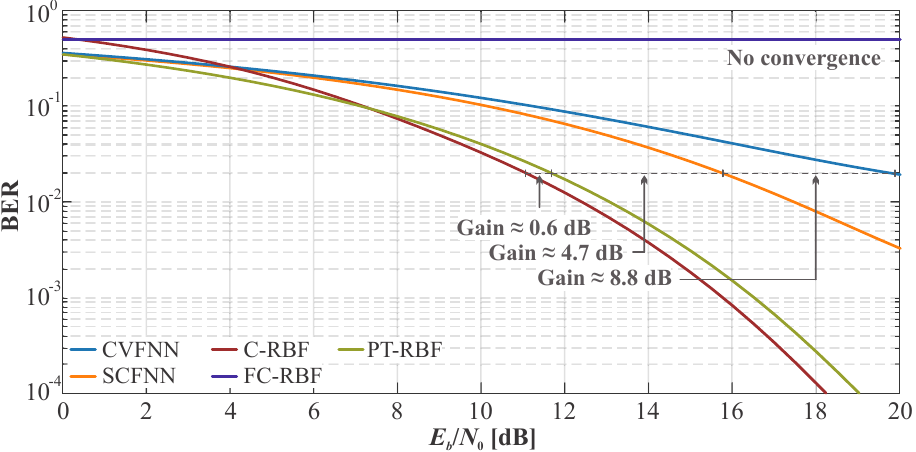}
\caption{BER inference results of the CVNNs for joint massive MIMO channel estimation and decoding. Results were averaged over ten subsequent simulations.}
\label{fig:mimo_ber}
\end{figure}

Fig.~\ref{fig:mimo_ber} presents the bit error rate (BER) versus $E_b/N_0$ inference results of the CVNNs averaged over ten subsequent simulations. The BER is computed after the CVNNs reach the steady-state MSE after 360 OFDM blocks (see Fig.~\ref{fig:mimo_mse}). As the FC-RBF cannot converge for $E_b/N_0\leq 20$~dB, its BER curve is static ($\mathrm{BER}=0.5$) for the whole range of $E_b/N_0$. For the sake of comparison, considering a $\mathrm{BER}=2\times 10^{-2}$, the C-RBF achieved the best performance, surpassing the PT-RBF, SCFNN, and CVFNN in 0.6~dB, 4.7~dB, and 8.8~dB, respectively. Unlike the previous benchmark results, the PT-RBF only presented the second-best results.

\begin{figure}[t!]
\centering
\includegraphics[width=\linewidth]{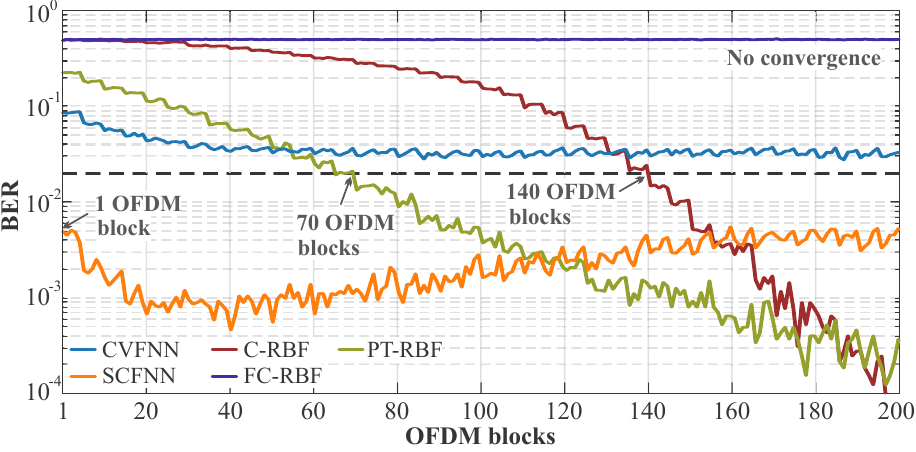}
\caption{Convergence rate results of the CVNNs for joint massive MIMO channel estimation and decoding at $E_b/N_0 = 20$~dB. SD-FEC BER threshold of $2\times 10^{-2}$.}
\label{fig:convergence_mimo_20dB}
\end{figure}

\begin{figure}[t!]
\centering
\includegraphics[width=\linewidth]{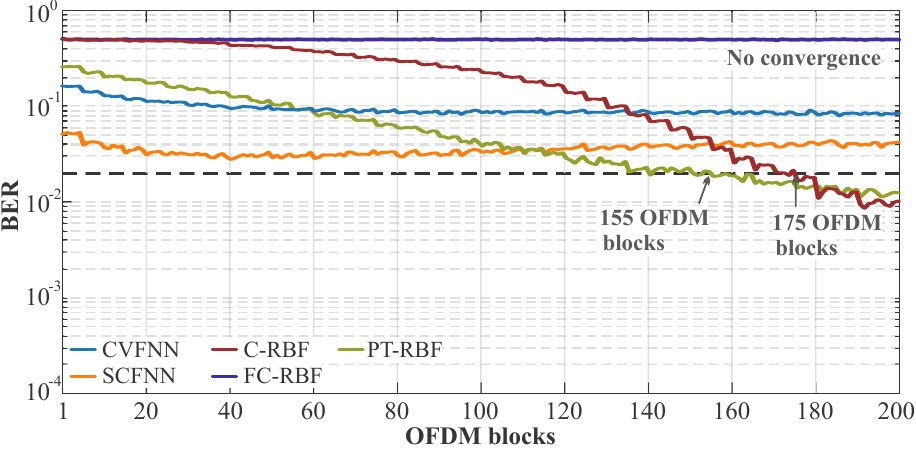}
\caption{Convergence rate results of the CVNNs for joint massive MIMO channel estimation and decoding at $E_b/N_0 = 14$~dB. SD-FEC BER threshold of $2\times 10^{-2}$.}
\label{fig:convergence_mimo_14dB}
\end{figure}

In our analysis, we utilized a soft-decision forward error correction (SD-FEC) BER threshold of $2 \times 10^{-2}$ to evaluate the performance of various CVNNs based on their convergence rates at different energy per bit to noise power spectral density ratio ($E_b/N_0$), specifically for $E_b/N_0 = 20$ dB and $E_b/N_0 = 14$ dB, as illustrated in Figs. \ref{fig:convergence_mimo_20dB} and \ref{fig:convergence_mimo_14dB}, respectively.

\begin{figure}[t!]
\includegraphics[width=\linewidth]{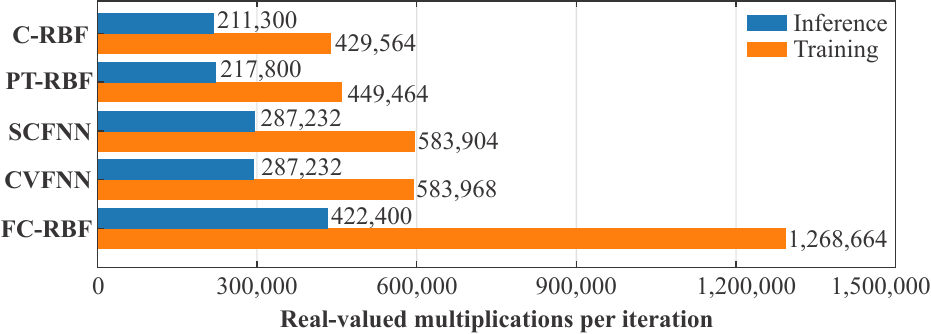}
\caption{Computational complexities of the CVNNs for joint channel estimation and decoding of massive MIMO communication systems.}
\label{fig:cvnns_mimo_comp}
\end{figure}

At an $E_b/N_0 = 20$ dB, both the FC-RBF and the CVFNN did not achieve the desired BER threshold. The C-RBF and PT-RBF models, which exhibited superior steady-state MSE performance, surpassed the SD-FEC BER threshold after processing 140 and 70 OFDM blocks, respectively. In contrast, the SCFNN rapidly met the threshold after just one pilot block. However, it was already demonstrated in the benchmark problems that the SCFNN is more susceptible to noise than the RBF-based CVNNs \cite{mayer2022b}.

When the noise is increased, for the case of $E_b/N_0 = 14$ dB, only the C-RBF and PT-RBF models crossed the SD-FEC BER threshold. The PT-RBF model achieved the threshold 20 OFDM blocks before the C-RBF, demonstrating the best convergence rate in noisy conditions.

Based on the computational complexities in \cite{mayer2022b} (see Tables 3.2, 3.3, 3.5, 3.6, and 3.7), Fig. \ref{fig:cvnns_mimo_comp} presents, for the CVNNs implemented for the massive MIMO problem, the real-valued multiplications per iteration of joint channel estimation and decoding, detailed on the right side of the horizontal bars. Training and inference complexities are depicted as orange and blue horizontal bars, respectively. The CVNNs were ordered by their training complexity.

The C-RBF and PT-RBF models not only delivered superior performance compared to the SCFNN and CVFNN but also exhibited lower computational complexities, making them ideal for complex and noisy environments. On the other hand, the FC-RBF, despite being an RBF-based model, failed to converge for $E_b/N_0 \leq 20$ dB and had more than double the computational complexity of the C-RBF. The limitations of the FC-RBF have been previously discussed in the literature. The FC-RBF noise dependence was addressed by Savitha et al.~\cite{Savitha2012}, in which the authors proposed metacognitive learning to filter the input data, regulating the learning process. However, this metacognitive learning technique is not interesting for time-variant systems because of the higher computational complexity.

\section{Conclusion}
This study introduces a comprehensive analysis of joint channel estimation and MIMO decoding in massive MIMO-OFDM systems utilizing various complex-valued neural networks (CVNNs) architectures with online training. The performance evaluation of CVFNN, SCFNN, C-RBF, FC-RBF, and PT-RBF architectures, in realistic 5G communication scenarios, demonstrates significant enhancements in both mean squared error (MSE) convergence and bit error rate (BER) accuracy. Particularly, the C-RBF and PT-RBF architectures show superior results, indicating their potential as robust and efficient solutions for complex and noisy communication environments. These findings suggest that RBF-based CVNNs can play a crucial role in advancing next-generation wireless systems. Future work will further refine these neural network models (e.g., with distinct initialization schemes, modified regularization techniques, and mini-batch learning) and explore different parameter configurations (e.g., modulation order and number of antennas). Moreover, CVNN applications in other challenging communication scenarios will be further investigated.

\section*{Acknowledgments}
This work was supported in part by the Coordenação de Aperfeiçoamento de Pessoal de Nível Superior --- Brazil
(CAPES) --- Finance Code 001.

\bibliographystyle{IEEEtran}
\bibliography{references.bib}
\balance

\end{document}